\documentclass{aa}

\usepackage{graphicx}
\usepackage{txfonts}
\usepackage{natbib}
\usepackage{color}
\usepackage{multirow} 	
\usepackage{relsize}		
									
\usepackage[normalem]{ulem}

\def\sss{\scriptscriptstyle}
\def\U{{\sss \!\mathrm{U}}}
\def\L{{\sss \!\mathrm{L}}}
\def\K{{\sss \!\mathrm{K}}}

\def\nur{\nu_\mathrm{r}}
\def\nuv{\nu_\theta}
\def\nuL{\nu_\L}
\def\nuU{\nu_\U}
\def\nuK{\nu_\K}

\def\asign#1#2{a^{*}_{\mathrm{#1:#2}}}
\def\amax#1#2{a_{\mathrm{#1:#2}}}
\def\amaxkep#1#2{\tilde{a}_{\mathrm{#1:#2}}}

\begin{document}

%-------------------------------------------------------------------------------
%                                     TITLE
%-------------------------------------------------------------------------------
\title{Super-spinning compact objects and models of high-frequency quasi-periodic oscillations observed in Galactic microquasars}
\subtitle{II.~Forced resonances}
%-------------------------------------------------------------------------------

\author
{
A.~Kotrlov\'a\inst{\ref{CCP}}\and E.~\v{S}r\'amkov\'a\inst{\ref{CCP}}\and G.~T\"or\"ok\inst{\ref{CCP}}\and Z.~Stuchl\'{\i}k\inst{\ref{CTP}}\and K.~Goluchov\'a\inst{\ref{CCP}\and\ref{CTP}\and\ref{CAMK}}
}

\institute{
Institute of Physics and Research Centre for Computational Physics and Data Processing,\\
Faculty of Philosophy \& Science, Silesian University in Opava, Bezru\v{c}ovo n\'am.~13, CZ-746\,01 Opava, Czech Republic\\
\email{Andrea.Kotrlova@fpf.slu.cz}\label{CCP}
\and
Institute of Physics and Research Centre for Theoretical Physics and Astrophysics,\\Faculty of Philosophy \& Science, Silesian University in Opava, Bezru\v{c}ovo n\'am.~13, CZ-746\,01 Opava, Czech Republic\label{CTP}
\and
Nicolaus Copernicus Astronomical Centre, Bartycka 18, PL-00716 Warsaw, Poland\label{CAMK}
}

\date{Received / Accepted}

\authorrunning{A. Kotrlov\'a et al.}
\titlerunning{Superspinars and resonance models of HF QPOs}

\date{Received / Accepted}

%-------------------------------------------------------------------------------
\abstract
{
In our previous work (Paper~I) we applied several models of high-frequency quasi-periodic oscillations (HF QPOs) to estimate the~spin of the~central compact object in three Galactic microquasars assuming the~possibility that the~central compact body is a~super-spinning object (or a~naked singularity) with external spacetime described by Kerr geometry with a~dimensionless spin parameter $a\equiv \mathrm{c}J/\mathrm{G}M^2>1$. Here we extend our consideration, and in a~consistent way investigate implications of a~set of ten resonance models so far discussed only in the~context of $a<1$. The~same physical arguments as in Paper~I are applied to these models, i.e. only a~small deviation of the~spin estimate from $a=1$, $a\gtrsim 1$, is assumed for a~favoured model. For five of these models that involve Keplerian and radial epicyclic oscillations we find the~existence of a~unique specific QPO excitation radius. Consequently, there is a~simple behaviour of dimensionless frequency $M\times\nuU (a)$ represented by a~single continuous function having solely one maximum close to $a\gtrsim1$. Only one of these models is compatible with the~expectation of $a\gtrsim 1$. The~other five models that involve the~radial and vertical epicyclic oscillations imply the~existence of multiple resonant radii. This signifies a~more complicated behaviour of $M\times\nuU (a)$ that cannot be represented by single functions. Each of these five models is compatible with the~expectation of $a\gtrsim 1$.
}
%-------------------------------------------------------------------------------
\maketitle

%-------------------------------------------------------------------------------
\section{Introduction}
\label{section:introduction}
%-------------------------------------------------------------------------------

Accreting black holes (BHs) or  neutron stars (NSs) are believed to constitute the~compact component in several tens of X-ray binaries. The~accretion disc contributes significantly to the~high X-ray luminosity of these objects. Most of the~X-ray radiation comes from the~inner parts of the~disc. Both the~BH and NS sources exhibit a~variability over a~wide range of frequencies. In addition to broad noise, their power density spectra (PDS) also contain relatively coherent features known as quasi-periodic oscillations (QPOs). A~brief introduction to the~subject of QPOs can be found in \citet{kli:2006} and \citet{mcc-rem:2006}.

%-------------------------------------------------------------------------------
%                          FIGURE
%-------------------------------------------------------------------------------
\begin{figure*}[t]
\begin{center}
a)\hfill ~~~b) \hfill $\phantom{c)}$
\includegraphics[width=1\hsize]{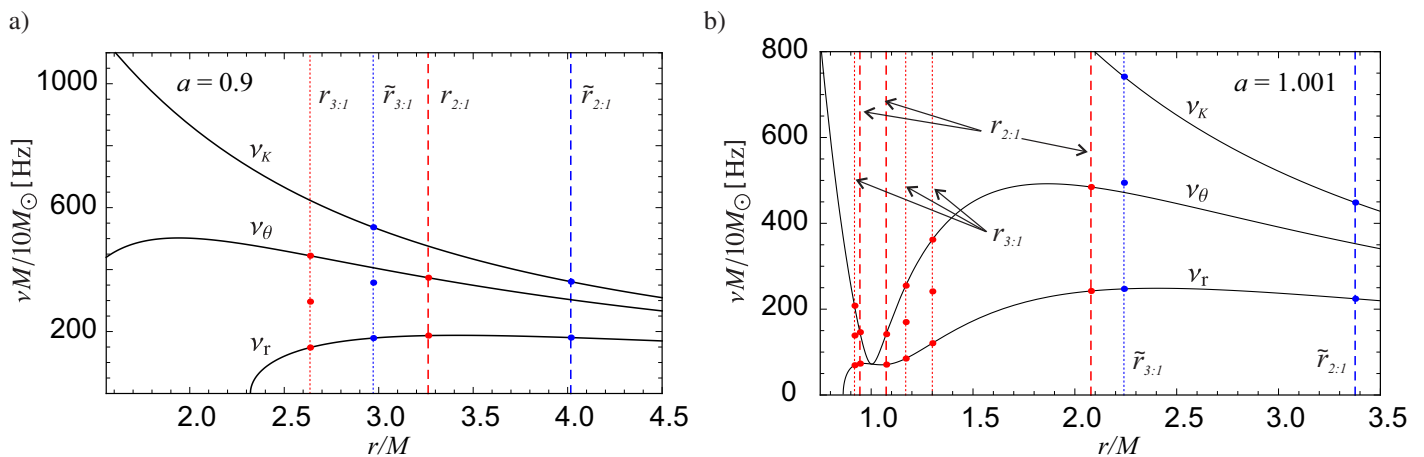}
\end{center}
\caption{a) Behaviour of orbital frequencies and location of several resonant radii determined by the~Ep and Kep models for $a=0.9$. b) Behaviour of orbital frequencies and location of several resonant radii determined by the~Ep and Kep models for $a=1.001$. For Kerr black holes there is only one specific resonant radius for each displayed resonance, while even three resonant radii may appear for a~single Ep model in the~case of a~super-spinning compact object or naked singularity.}
\label{figure:orbits}
\end{figure*}
%-------------------------------------------------------------------------------
%
%-------------------------------------------------------------------------------

%\subsection{High-frequency QPOs observed in BH systems}

The~X-ray PDS of several stellar mass BH systems display low-frequency (LF) QPOs in the~range of 0.1--30\,Hz. Their strength expressed in terms of the~fractional root-mean-squared (rms) amplitude $\mathcal{R}$ is sometimes as high as $\mathcal{R}\sim20\,\%$. %\footnote{In other words, fraction of the~mean count rate related to the~QPO can be close to 20\%.}
Much attention among theoreticians is, however, attracted to high-frequency (HF or kHz) QPOs that display their frequencies within the~range of 40--450\,Hz. High-frequency QPOs usually display much smaller amplitudes than LF QPOs, $\mathcal{R}\sim 1-4\,\%$, but their frequencies correspond to orbital timescales in the~vicinity of a~BH. It is believed that this coincidence is a~strong indication that the~corresponding signal originates in the~innermost parts of the~accretion disc. This belief has also been supported by means of the~Fourier-resolved spectroscopy \citep[e.g.][]{gilf-etal:2000}.

The~kHz QPO peaks are detected at rather constant frequencies characteristic for a~given source. Moreover, it has been noticed \citep{abr-klu:2001, mcc-rem:2006} that they usually appear in ratios of small natural numbers, typically in the~$3\!:\!2$ ratio. Discussion of these measurements and their relation to other sources can be found in \citet{mcc-rem:2006}, \citet{tor-etal:2005:AA:}, \citet{tor:2005:AA:}, and \citet{kli:2006}. It has been noticed that the~{3:2} frequencies observed in the~Galactic microquasars are well matched by the~relation \citep[e.g.][]{mcc-rem:2006}
%---------------
\begin{equation}
%---------------
\label{equation:bestfit}
\nuL= 1.862 \left ({\frac{M}{\mathrm{M}_{\odot}}}\right)^{-1}\mathrm{kHz},
%---------------
\end{equation}
%---------------
where $\nuL$ is the~lower of the~two frequencies forming the~3:2 ratio,
%---------------
\begin{equation}
%---------------
R=\nuU/\nuL= 3/2\,.
%---------------
\end{equation}
%---------------
\noindent
This scaling of QPO frequencies further supports the~hypothesis of their orbital origin \citep[e.g.][]{tor-etal:2005:AA:}.

%\subsection{Properties of compact objects}

There is a~common conviction in the~astrophysical community that studying X-ray spectra and variability can be used to put constraints on the~properties of compact objects such as BH mass $M$ and spin {$a \equiv \mathrm{c}J/\mathrm{G}M^2$}. A~standard way to measure the~BH spin is based on different spectral fitting methods, namely by fitting the~X-ray spectral continuum or the~relativistically broadened Fe K alpha lines. Using these spectral fitting methods, many authors have carried out estimations of BH spin \citep[see e.g.][]{mcc-rem:2006,mid-etal:2006,don-etal:2007,mcc-etal:2008,mil:2007,sha-etal:2008,mcc-etal:2010,mcc-etal:2011,mcc-etal:2014}.

\citet{abr-etal:2004:ApJ:} have suggested that detections of 3:2 QPOs could provide a~basis for another method for the~determination of masses of BH sources such as the~active galactic nuclei and ultraluminuos X-ray sources. This suggestion is further supported by \citet{tor:2005:AA:,tor:2005:AN:} who has discussed the~possibility of orbital resonance origin of QPOs in the~Galactic centre BH Sagittarius~A$^*$ measured by \citet{asc-etal:2004}. It has been argued that the~confirmation of 3:2 QPOs in this source could be of fundamental importance to the~BH accretion theory. At present, a~full decade after these results were published and even though Sgr~A$^*$ QPO detections remain questionable, there is a~growing evidence of the~existence of a~large-scale validity of the~$1/M$ scaling of the~BH QPO frequencies (\ref{equation:bestfit}) \citep[see e.g.][]{zho-etal:2015}.

%\subsection{Determination of spin from HF QPOs}

For a~BH candidate which displays HF QPOs and whose mass has been measured by independent measurements, the~BH spin estimate can in principle be obtained by simply comparing the~observed QPO frequencies to QPO frequencies predicted by a~particular QPO model. Through this approach, using miscellaneous QPO models, a~number of authors have attempted to estimate the~spin of the~central Kerr BH in three Galactic microquasars -- GRS~1915$+$105, XTE~J1550$-$564, and GRO~J1655$-$40 -- that display the~3:2 twin peak QPOs and have known masses \citep[e.g.][]{abr-klu:2001,wag-etal:2001,kat:2004,tor:2005:AN:,wag:2012,hor-don:2014,ort-etal:2014,mot-etal:2014,stu-kol:2016:ApJ,stu-kol:2016:AA}.

This approach can also be applied in the~context of testing hypotheses alternative to consideration of Kerr BHs \citep[][]{tor-stu:2005:AA:,kot-etal:2008:CQG:,psa-etal:2008,bam-fre:2009,gim-hor:2009,stu-kot:2009,stu-sch:2010,stu-sch:2012b,stu-sch:2012a,stu-sch:2013,bam:2011,joh-psa:2011,li-bam:2013b,li-bam:2013a,bam:2012,bam:2014,kol-stu:2013,yag-ste:2016:,joh:2016}. Here we focus on the~QPO based spin estimates of compact object properties, in particular on those assuming orbital resonances in Kerr spacetimes.

%--------------------------------------------------------------------
%          TABLE: MODELS
%--------------------------------------------------------------------
\begin{table*}[t]
\caption{Frequency relations corresponding to individual epicyclic and Keplerian forced resonance models.}
\label{table:models}
\renewcommand{\arraystretch}{1.4}
\begin{center}
\begin{tabular}{llllc}
    \hline
  \hline
     \textbf{Model} & \textbf{Type of resonance} & \multicolumn{2}{c}{\textbf{Relations}} & $\nuv/\nur$\,\ \textbf{or~}\,$\nuK/\nur$ \\
    \hline \hline
\textbf{Ep1} & epicyclic 3:1 forced & $\nuL=\nuv-\nur$ & $\nuU=\nuv$ & $3/1\phantom{^*}$ \\
\textbf{Kep1} & Keplerian 3:1 forced & $\nuL=\nuK-\nur$ & $\nuU=\nuK$ & $3/1\phantom{^*}$ \\
\hline
\textbf{Ep2} & epicyclic 2:1 forced & $\nuL=\nuv$ & $\nuU=\nuv+\nur$ & $2/1\phantom{^*}$ \\
\textbf{Kep2} & Keplerian 2:1 forced & $\nuL=\nuK$ & $\nuU=\nuK+\nur$ & $2/1\phantom{^*}$ \\
\hline
\textbf{Ep3} & epicyclic 5:1 forced & $\nuL=\nuv-\nur$ & $\nuU=\nuv+\nur$ & $5/1\phantom{^*}$ \\
\textbf{Kep3} & Keplerian 5:1 forced & $\nuL=\nuK-\nur$ & $\nuU=\nuK+\nur$ & $5/1\phantom{^*}$ \\
\hline
\textbf{Ep4} & epicyclic 5:2 forced & $\nuL=\nur$ & $\nuU=\nuv-\nur$ & $5/2\phantom{^*}$ \\
\textbf{Kep4} & Keplerian 5:2 forced & $\nuL=\nur$ & $\nuU=\nuK-\nur$ & $5/2\phantom{^*}$ \\
\hline
\textbf{Ep5} & epicyclic 5:3 forced & $\nuL=\nuv-\nur$ & $\nuU=\nur$ & $5/3\phantom{^*}$ \\
\textbf{Kep5} & Keplerian 5:3 forced & $\nuL=\nuK-\nur$ & $\nuU=\nur$ & $5/3\phantom{^*}$ \\
\hline \hline
\end{tabular}
\end{center}
\end{table*}

%------------------------------------------------------------------------------
\section{Estimates of spin in Galactic microquasars and forced resonances}
%------------------------------------------------------------------------------

%\subsection{Consideration of individual QPO models}

Orbital resonances have been proposed as a~generic QPO mechanism in the~works of \citet{abr-klu:2001} and \citet{klu-abr:2001}, and previously discussed in the~general framework of orbital motion by \citet{ali-gal:1981}.  Several consequent studies have elaborated this proposal \citep[e.g.][]{abr-etal:2002:,klu-abr:2002:,klu-abr:2005:,abr-etal:2003b,abr-etal:2003c,reb:2004,hor:2004,hor:2005b,sra:2005:,hor-kar:2006,vio-etal:2006,sra-etal:2007,hor:2008,reb:2008,stu-etal:2008b, hor-etal:2009, stu-etal:2013:AA:,stu-etal:2014:ACTA:}.\footnote{These works often assume the~formation of an inner accretion torus, which is frequently considered in the~context of the~generic QPO mechanism; see also \citet{rez-etal:2003,zan-etal:2005:,sch-rez:2006:,Montero-etal:2007:,Ing-Don:2009:,Ing-Don:2010:,Ing-Don:2011:,Ing-Don:2016:,tor-etal:2016:MNRAS:} and references therein.}
In the~works of \citet{klu-abr:2001} and \citet{tor-etal:2005:AA:}, spin estimates have been carried out for the~three microquasars based on a~group of QPO models that deal with a~non-linear resonance between some modes of accretion disc oscillations. In particular, \citet{tor-etal:2005:AA:} have discussed resonances that occur between axisymmetric radial and vertical disc oscillations and that may also involve Keplerian motion.

Assuming fluid-disc-oscillation frequencies, it is expected that resonances occur at preferred resonant orbits (QPO excitation radii), $r_\mathrm{p\,:\,q}$ or $\tilde{r}_\mathrm{p\,:\,q}$, where the~vertical and radial epicyclic frequency, $\nur$ and $\nuv$, or the~Keplerian ($\nuK$) and radial epicyclic frequency are in ratios of small rational numbers,
%----------------------
\begin{eqnarray}
\nuv(r_\mathrm{p\,:\,q})=\left(\frac{p}{q}\right)\nur(r_\mathrm{p\,:\,q})\,,\\
\nuK(\tilde{r}_\mathrm{p\,:\,q})=\left(\frac{p}{q}\right)\nur(\tilde{r}_\mathrm{p\,:\,q})\,.
\end{eqnarray}
%----------------------
The~3:2 ratio of the~observed frequencies then arises either from the~generic 3:2 ratio of the~resonant eigenfrequencies or due to resonant combinational frequencies. The~well-known explicit formulae for epicyclic frequencies in Kerr spacetimes have been discussed in several works \citep[e.g.][]{ali-gal:1981,now-leh:1999,tor-stu:2005:AA:}. The~behaviour of orbital frequencies and the~positions of several specific resonant orbits are illustrated in Fig.~\ref{figure:orbits} assuming two different values of $a$ corresponding to BH and non-BH Kerr spacetimes. For a~fixed $a$, these frequencies scale with the~compact object mass $M$ as $\nu \propto 1/M$.

\subsection*{Consideration of Kerr BHs and superspinars (or naked singularities) and resonance QPO models}

Several variations of resonance models have been assumed by \citet{tor-etal:2005:AA:}. These consist of both parametric and forced non-linear resonances, in particular the~3:2 parametric resonance, and the~3:1, 2:1, 5:1, 5:2, 5:3 forced resonances. The~two most popular alternatives
%of models have been under the~consideration. These are represented by
are under consideration: the~epicyclic resonance (Ep) model dealing with radial and vertical epicyclic oscillations and the~Keplerian resonance (Kep) model assuming a~resonance between the~orbital Keplerian motion and the~radial epicyclic oscillations.

Further complex analysis of spin estimations has been done by \citet{tor-etal:2011:AA:} for a~set of various QPO models. It has been performed namely for hot-spot-like models, models assuming a~certain form of resonance between some accretion disc oscillatory modes, and for discoseismic models. Individual models considered in \citet{tor-etal:2005:AA:} and  \citet{tor-etal:2011:AA:} point to very different values of $a$ covering the~whole BH range $a\in[0,1]$. The~iron-line profile, X-ray continuum, and the~QPO based spin estimates reveal good agreement only in certain particular applications \citep[e.g.][]{ste-etal:2011}. There are still discrepancies, and no agreement between the~three approaches has been achieved so far \citep[see e.g.][]{tor-etal:2011:AA:,kot-etal:2014:AA:}.

In \citet{kot-etal:2014:AA:} (hereafter Paper~I) we extended the~consideration of QPO based spin estimates to the~non-BH case and investigated the~behaviour of  $M\times\nuU (a)$ relations implied by individual QPO models for $a>1$. We discussed various types of models including several resonance models and concluded that the~Ep and Kep models are favoured in the~context of the~$a>1$ hypothesis. In Paper~I, our consideration did not include forced resonances investigated by \citet{tor-etal:2005:AA:} for Kerr BHs. Here we explore the~predictions of ten of the forced resonance models mentioned above. The~list of these models together with associated relations giving observable QPO frequencies is given in Table~\ref{table:models}.

%--------------------------------------------------------------------
%\section{Results for the Keplerian models}
\section{Keplerian models}
%--------------------------------------------------------------------

%-------------------------------------------------------------------
%-------------------------------------------------------------------------------
%                          FIGURE
%-------------------------------------------------------------------------------
\begin{figure*}[t]
\begin{center}
a)\hfill ~~~b) \hfill $\phantom{c)}$
\includegraphics[width=1\hsize]{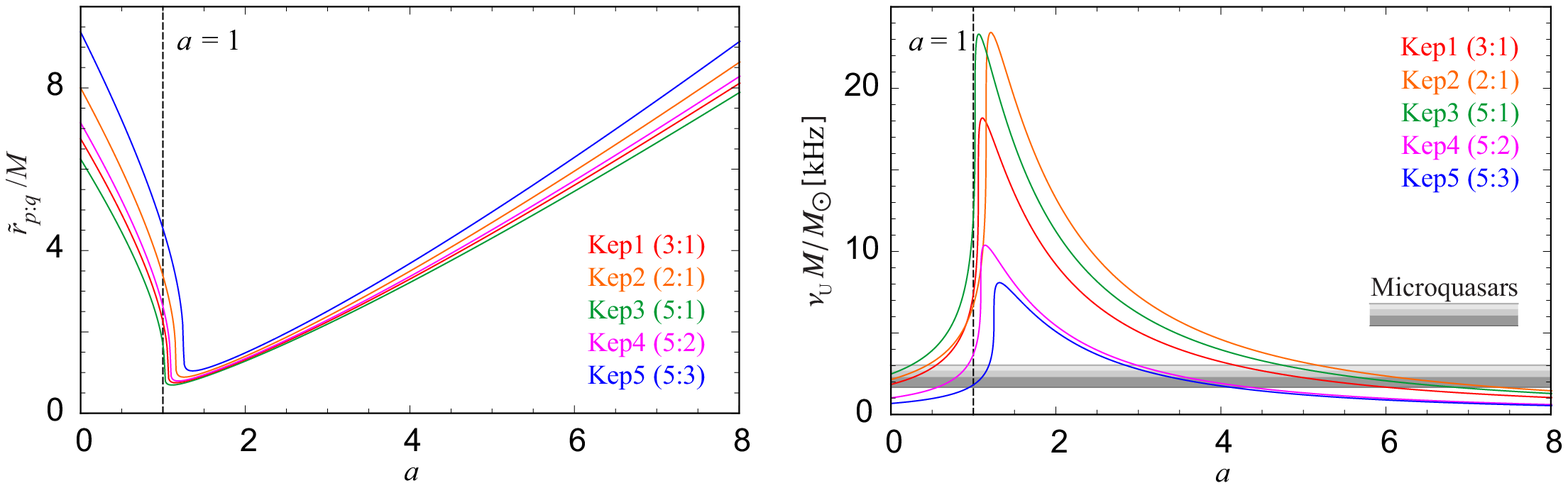}
\end{center}
\caption{a) Resonant radii of Kep models calculated in Kerr spacetimes. b) Resonant frequencies of Kep models calculated in Kerr spacetimes. The~shaded region corresponds to observational values of $\nuU\times M$ determined for Galactic microquasars (see Fig.~\ref{figure:nuEpicyclic} for details and Sect.~\ref{section:conclusions} for a~discussion).}
\label{figure:Keplerian}
%\end{figure*}
%-------------------------------------------------------------------------------
%
%-------------------------------------------------------------------------------

%--------------------------------------------------------------------

%-------------------------------------------------------------------------------
%                          FIGURE
%-------------------------------------------------------------------------------
\bigskip

\begin{center}
a)\hfill ~~~b) \hfill $\phantom{c)}$
\includegraphics[width=1\hsize]{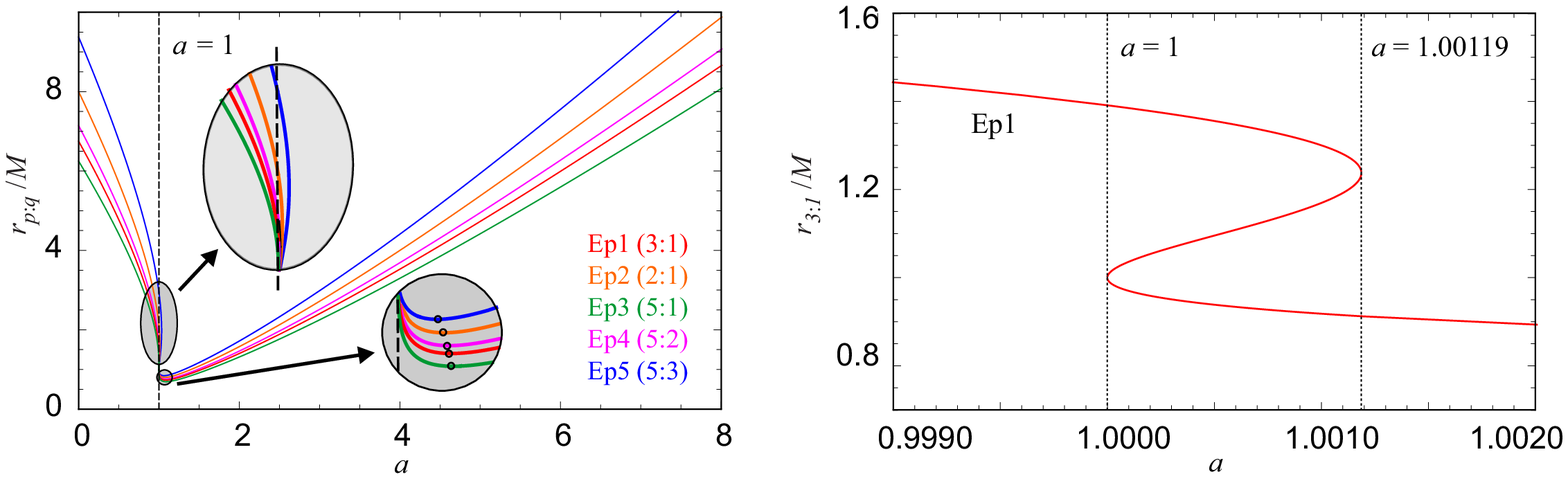}
\end{center}
\caption{a) Resonant radii of Ep models calculated in Kerr spacetimes. The~enlarged area emphasizes a~region close to $a=1$ where each {$a(r_\mathrm{p:q})$} function exhibits both a~local minimum and a~local maximum. Global minima of ${r}_\mathrm{p:q}(a)$ functions are denoted by circles. b) Detailed view of the~behaviour of the~{$r_{\mathrm{3:1}}(a)$} relation for $a\approx1$. {The~global minimum of $r_{3:1}$ is not shown since it occurs only for a~higher value of spin, $\amax{3}{1}=1.08471$.}}
\label{figure:Epicyclic}
\end{figure*}
%-------------------------------------------------------------------------------
%
%-------------------------------------------------------------------------------

In this section we examine the~predictions of the~forced resonance models assuming Keplerian and radial epicyclic oscillations. Paper~I illustrates that the~3:2 parametric Kep model implies a~much simpler behaviour of the~$M\times\nuU (a)$ relation than the~3:2 parametric Ep model. This is also true for the~forced resonance models. As noted by \citet{tor-stu:2005:AA:}, the~ratio between Keplerian frequency and radial (or vertical) epicyclic frequency is a~monotonic function of orbital radius $r$ for any value of $a\in[0,\,\infty)$. For this reason, there is always a~unique QPO excitation radius $\tilde{r}_{\mathrm{p:q}}$ for any Kep model.

In the~particular case of the~3:1 forced Keplerian resonance model (Kep1 model), the~resonance occurs at the~$\tilde{r}_{\mathrm{3:1}}$ orbit where
%-----------------
\begin{equation}
\label{equation:3r:1K}
\nuK=3\nur\,
\end{equation}
%-----------------
and we can express the~observed 3:2 QPO frequencies in the~same manner as \citet{klu-abr:2001},
%-----------------
\begin{equation}
\label{equation:3r:1K:frequencies}
\nuL=\nuK-\nur,~\mathrm{and}~ \nuU=\nuK\,.
\end{equation}
%-----------------

Solving Equation (\ref{equation:3r:1K}) yields the~relation between spin and dimensionless resonant radius $\tilde{x}_{\mathrm{3:1}}\equiv \tilde{r}_{\mathrm{3:1}}/M$,
%-----------------
\begin{equation}
\label{equation:3r:1K:spin}
a=\frac{\sqrt{\tilde{x}_{\mathrm{3:1}}}}{9}\left(12 \mp \sqrt{6(4\tilde{x}_{\mathrm{3:1}}-3)}\right) \,,
\end{equation}
%-----------------
where the~`{$-$}' sign holds for {$a\leq\amaxkep{3}{1}=1.1547$}. The~associated $\tilde{x}_{\mathrm{3:1}}(a)$ function is shown in Fig.~\ref{figure:Keplerian}a. Figure~\ref{figure:Keplerian}b then shows the~related $M\times\nuU (a)$ function exhibiting its maximum for {$1\lesssim a\approx\amaxkep{3}{1}$}.

The~other four Kep models (Kep2--Kep5) exhibit qualitatively similar behaviour of resonant radii to that of the~Kep1 model. In full analogy, $M\times\nuU (a)$  relations are then also represented by single continuous functions having a~maximum close to $1\lesssim a\approx\amaxkep{p}{q}$. In Table~\ref{table:spin_functionI} we specify relations between the~spin and resonant radii calculated for each of the~Kep1--Kep5 models. Illustrations of relevant $M\times\nuU (a)$  relations are included in Fig.~\ref{figure:Keplerian}.

%-------------------------------------------------------------------------------
%                          FIGURE
%-------------------------------------------------------------------------------
\begin{figure*}[t]
\begin{center}
a)\hfill ~~~~~b) \hfill $\phantom{c)}$
\includegraphics[width=1\hsize]{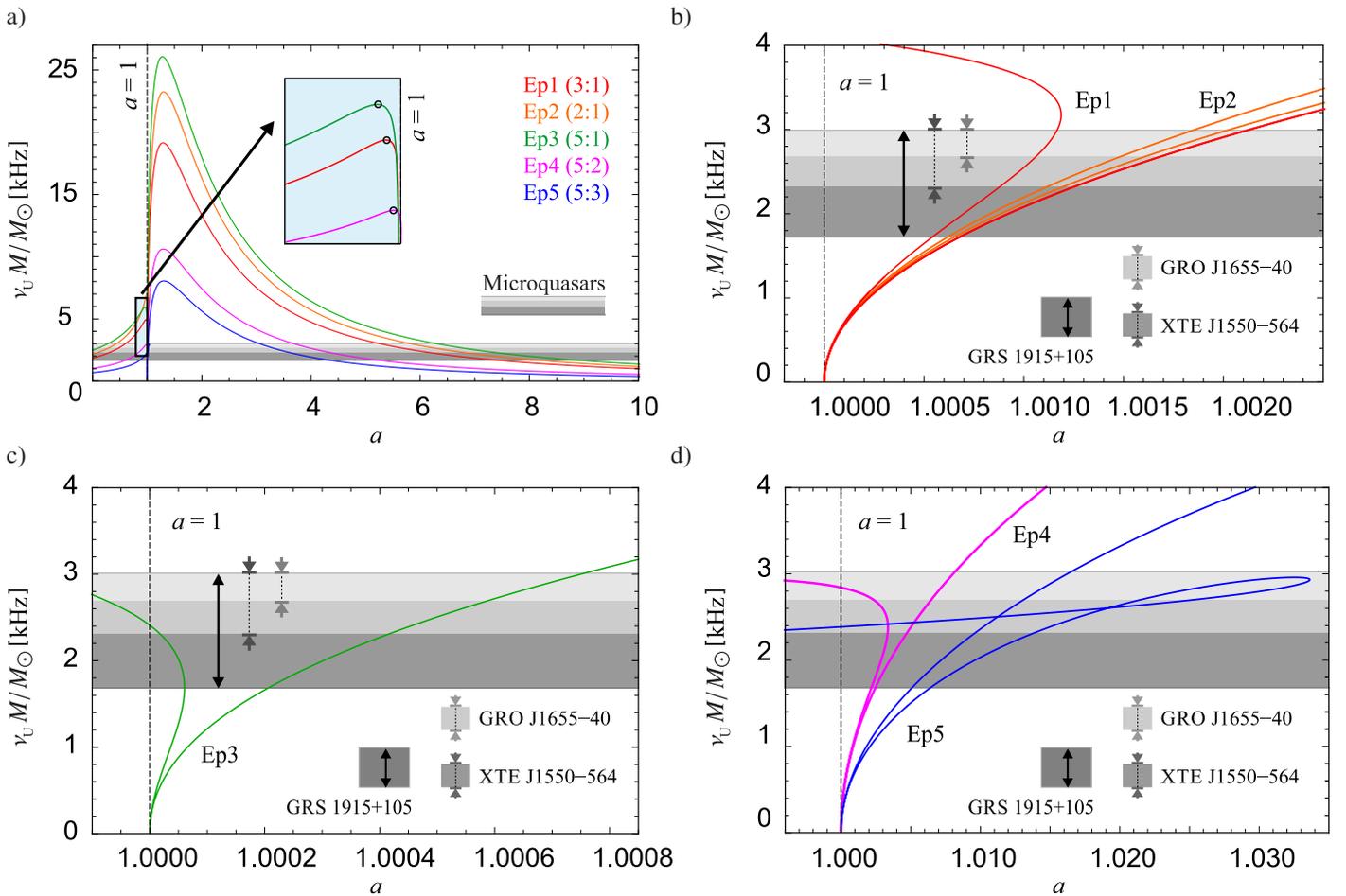}
\vspace{-43ex}

c)\hfill ~~~~~d) \hfill $\phantom{e)}$

\vspace{38ex}

\end{center}
\caption{Resonant frequencies of the~Ep models calculated in Kerr spacetimes. The~enlarged area emphasizes the~existence of maxima of the~$\nuU\times M$ curves in the~case of the~Ep1, Ep3, and Ep4 models in BH spacetimes. The~shaded region corresponds to the~observational values of $\nuU\times M$ determined for Galactic microquasars (see Sect.~\ref{section:conclusions} for discussion). a) Behaviour of resonant frequencies of the~whole group of models on a~large scale of $a$. b) Detailed view of behaviour of the~resonant frequencies for the~Ep1 and Ep2 models for $a\approx1$. The~individual horizontal grey-scaled areas denote the~observational values of $\nuU\times M$ determined for each of the~Galactic microquasars. c) Detailed view of the~behaviour of resonant frequencies for the~Ep3 model for $a\approx1$. d) Detailed view of the~behaviour of resonant frequencies for the~Ep4 and Ep5 models for $a\approx1$.}
\label{figure:nuEpicyclic}
\end{figure*}
%-------------------------------------------------------------------------------
%
%-------------------------------------------------------------------------------

%--------------------------------------
%\section{Results for the epicyclic models}
\section{Epicyclic models}
%--------------------------------------
\label{section:topologies}

In analogy to Equation (\ref{equation:3r:1K}), for the~3:1 forced epicyclic resonance model (Ep1 model), the~resonance occurs at the~$r_{3:1}$ orbit where we have
%-----------------
\begin{equation}
\label{equation:3r:1v}
\nuv=3\nur\,
\end{equation}
%-----------------
and we can express the~observed 3:2 QPO frequencies as \citet{klu-abr:2001},
%-----------------
\begin{equation}
\label{equation:3r:1v:frequencies}
\nuL=\nuv-\nur~\mathrm{and}~ \nuU=\nuv\,.
\end{equation}
%-----------------
Solving Equation (\ref{equation:3r:1v}) yields the~relation between the~spin and dimensionless resonant radius ${x}_{\mathrm{3:1}}$,
%-----------------
\begin{equation}
\label{equation:3r:1v:spin}
a = \frac{\sqrt{x_{3:1}}}{15} \left(19\mp 2 \sqrt{15x_{3:1}-11}\right),
\end{equation}
%-----------------
where the~`{$-$}' sign holds for $a\leq {a}_{\mathrm{3:1}} = 1.08471$.

The~behaviour of relation (\ref{equation:3r:1v:spin}) is shown in Fig. \ref{figure:Epicyclic}a together with relations obtained for the~other four Ep models. On a~large scale of $a$, the~displayed curves are similar to those drawn for the~Kep models in Fig.~\ref{figure:Keplerian}a. However, there is a~qualitative difference which is emphasized within the~enlarged area in Fig.~\ref{figure:Epicyclic}a. For $a\gtrsim1$, each of Ep models exhibits `{S-like}' behaviour of resonant radii. In contrast to Kep models, the~resonance radius is not a~single function of spin because of the~non-monotonic behaviour of the~ratio between epicyclic frequencies noted and discussed by \citet{tor-stu:2005:AA:} and \citet{stu-etal:2013:AA:}. We find that for Ep models and  $1\lesssim a \lesssim \asign{p}{q}$ three distinct branches of $x_\mathrm{p:q}(a)$ appear. This is illustrated in detail for the~Ep1 model in Fig.~\ref{figure:Epicyclic}b. {We specify the~relations between the~spin and resonant radii calculated for each of the~Ep1--Ep5 models in Table~\ref{table:spin_functionII}.}
% while the~number of branches $x_{p:q}(a)$ for any $a$ and each of Ep models can be found in Table~\ref{table:branches}.

The~behaviour of resonant frequencies $\nuU(a)$ implied by Ep models displayed in Fig.~\ref{figure:Epicyclic} is more complex than that implied by Kep models. In analogy to the~comparison between the~behaviours of resonant radii, on a~large scale of $a$, there is a~similarity between the~behaviours of frequencies predicted by both Kep and Ep models (see Figs.~\ref{figure:Keplerian}b and \ref{figure:nuEpicyclic}a). It can also be found that there is a~clear qualitative difference for $a\sim1$. Even for $a\lesssim1$, only the~Ep2 and Ep5 models imply qualitatively the~same monotonic increase in $\nuU$ with increasing $a$ as in the~Kep models. The~other three  Ep models (Ep1, Ep3, and Ep4) display sharp maxima of $\nuU(a)$ for rapidly rotating Kerr BHs (see the~enlarged area in Fig.~\ref{figure:nuEpicyclic}a). The~existence of these maxima is a~consequence of the~existence of the~maximum of the~vertical epicyclic frequency $\nuv(r)$ discussed by \citet{tor-stu:2005:AA:}. For {$1\lesssim a\lesssim \asign{p}{q}$}, the~$M\times\nuU(a)$ curves then have the~same topology for the~Ep1, Ep3, and Ep4 models, and two additional and different topologies for the~Ep2 and Ep5 models. These topologies are illustrated in Figs.~\ref{figure:nuEpicyclic}b--\ref{figure:nuEpicyclic}d.

%************************************************************************
%%%%%%%%%   Tabulka - spinova funkce I
%************************************************************************
\begin{table}[ht]
\caption{Spin functions $a(\tilde{x}_{\mathrm{p:q}})$ determined for the~Keplerian forced resonance QPO models. There are two separate branches of solution for each model, the~`{$-$}' sign applies to black holes and naked singularities with {$1<a\lesssim \amaxkep{p}{q}$}, while the~`{$+$}' sign applies to naked singularities with {$a \gtrsim \amaxkep{p}{q}$}. The~last column indicates the~value of $\amaxkep{p}{q}$ for each model.}
\renewcommand{\arraystretch}{1.8}
\label{table:spin_functionI}
  \begin{center}
\begin{tabular}{c|l|c}
  \hline
  \hline
  \textbf{Model} & \textbf{Spin function} & $\amaxkep{p}{q}$ \\
    \hline \hline
\textbf{Kep1} & $a(\tilde{x}_{3:1})=\frac{\sqrt{\tilde{x}_{3:1}}}{9}\left(12\mp\sqrt{6(4\tilde{x}_{3:1}-3)}\right)$ & $1.15470$  \\
%\hline
\textbf{Kep2} & $a(\tilde{x}_{2:1})=\frac{\sqrt{\tilde{x}_{2:1}}}{6}\left(8\mp\sqrt{9\tilde{x}_{2:1}-8}\right)$ & $1.25708$ \\
%\hline
\textbf{Kep3} & $a(\tilde{x}_{5:1})=\frac{\sqrt{\tilde{x}_{5:1}}}{15}\left(20\mp\sqrt{2 (36\tilde{x}_{5:1}-25)}\right)$ & $1.11111$  \\
%\hline
\textbf{Kep4} & $a(\tilde{x}_{5:2})=\frac{\sqrt{\tilde{x}_{5:2}}}{15}\left(20\mp\sqrt{63\tilde{x}_{5:2}-50)}\right)$ & $1.18783$  \\
%\hline
\textbf{Kep5} & $a(\tilde{x}_{5:3})=\frac{\sqrt{\tilde{x}_{5:3}}}{15}\left(20\mp\sqrt{2(24\tilde{x}_{5:3}-25)}\right)$ & $1.36083$  \\
\hline \hline
\end{tabular}
\end{center}
\end{table}

%------------------------------------------
%----------------------------------------------------

%************************************************************************
%%%%%%%%%   Tabulka - spinova funkce II
%************************************************************************
\begin{table}[ht]
\caption{Spin functions $a(x_{\mathrm{p:q}})$ determined for the~epicyclic forced resonance QPO models. There are two separate branches of solution for each model, the~`{$-$}' sign applies to black holes and naked singularities with {$1<a\lesssim \amax{p}{q}$}, while the~`{$+$}' sign applies to naked singularities with {$a \gtrsim \amax{p}{q}$}. The~last column indicates the~value of $\amax{p}{q}$ for each model as well as the~related spin value {$\asign{p}{q}$}, where the~$a(x_{\mathrm{p:q}})$ function has a~local maximum, which implies the~`{S-like}' behaviour of resonant radii $x_{\mathrm{p:q}}(a)$ and the~existence of three resonant radii for {$1 \lesssim a \lesssim \asign{p}{q}$} (see Fig.~\ref{figure:Epicyclic}b).}
\renewcommand{\arraystretch}{1.8}
\label{table:spin_functionII}
\begin{center}
\begin{tabular}{c|l|c}
    \hline
  \hline
     \textbf{Model} & \textbf{Spin function} & $\amax{p}{q}$ \\
		 & & $\asign{p}{q}$ \\
    \hline \hline
\textbf{Ep1} & $a(x_{3:1})=\frac{\sqrt{x_{3:1}}}{15} \left(19\mp2\sqrt{15x_{3:1}-11}\right)$ & $1.08471$ \\
& & $1.00119$ \\
\hline
\textbf{Ep2} & $a(x_{2:1})=\frac{\sqrt{x_{2:1}}}{5}\left(6\mp\sqrt{5x_{2:1}-4}\right)$ & $1.07331$ \\
& & $1.01193$ \\
\hline
\textbf{Ep3} & $a(x_{5:1})=\frac{\sqrt{x_{5:1}}}{13}\left(17\mp2\sqrt{13x_{5:1}-9}\right)$ & $1.08807$ \\
& & $1.00006$ \\
\hline
\textbf{Ep4} & $a(x_{5:2})=\frac{\sqrt{x_{5:2}}}{29}\left(36\mp\sqrt{7(29x_{5:2}-22)}\right)$ & $1.08123$ \\
& & $1.00337$ \\
\hline
\textbf{Ep5} & $a(x_{5:3})=\frac{\sqrt{x_{5:3}}}{51}\left(59\mp2 \sqrt{2 (51x_{5:3}-43)}\right)$ & $1.06226$ \\
& & $1.03363$ \\
\hline
  \hline
\end{tabular}
\end{center}
\end{table}
%-----------------------
%-----------------------

%------------------------------------------
\section{Discussion and conclusions}
%------------------------------------------
\label{section:conclusions}

As discussed in detail in Paper~I, numerous attempts to estimate the~spin $a$ of black hole candidates have been undertaken using the~iron-line profile, the~X-ray continuum, or the~QPO frequency fitting-methods. In Paper~I we attempted to extend the~application of the~X-ray timing approach to super-spinning compact objects with exteriors described by the~spacetime of a~Kerr naked singularity and interior given by the~solutions of string theory. We discussed there various types of QPO models including several resonance models and concluded that the~Ep and Kep models are favoured in the~context of the~$a>1$ hypothesis. Here we extended our consideration and investigated in a~consistent way the~implications of a~set of ten forced resonance models including five Kep and five Ep models so far discussed only in the~context of $a<1$. We applied the~same physical arguments to these models as used in Paper~I. In particular, we assumed that only a~small deviation of spin estimate from $a=1$, $a\gtrsim 1$ could occur for a~favoured model, as is often considered in the~case of super-spinning relativistic compact objects \citep[][see discussion in Paper~I]{cal-nob:1979,stu:1980,stu:1981,stu-etal:2011}.

%************************************************************************
%%%%%%%%%  Tabulka - vysledky pro vsechny tri mikrokvasary -- 3:2
%************************************************************************
\begin{table*}[t]
\caption{Spin intervals inferred for the~three microquasars (GRS~1915$+$105, XTE~J1550$-$564, and GRO~J1655$-$40) from the~individual Ep and Kep models.}
\label{table:spin}
\renewcommand{\arraystretch}{1.4}
\begin{center}
\begin{tabular}{c|lc|lc|lc}
\hline
\hline
\textbf{Model} & \multicolumn{2}{c}{\textbf{GRS~1915$+$105}} & \multicolumn{2}{|c|}{\textbf{XTE~J1550$-$564}} & \multicolumn{2}{c}{\textbf{GRO~J1655$-$40}}\\
\hline \hline
  \textbf{Ep1}  & $<0.61$             &                    & $0.32 - 0.59$        &               & $0.50 - 0.59$       &  \\
                & $1.00052 - 1.00217$ &  $5.25 - 7.40$     & $1.00088 - 1.00116$  &               & $1.00107 - 1.00116$ & \\
                &                     &                    & $1.00125 - 1.00210$  & $5.29 - 6.13$ & $1.00171 - 1.00209$ & $5.30 - 5.61$\\
\hline
  \textbf{Ep2}  & $<0.44$             &                    & $0.11 - 0.42$        &               & $0.31 - 0.42$       &  \\
                & $1.00059 - 1.00207$ & $5.81 - 8.14$      &  $1.00111 - 1.00200$ & $5.86 - 6.77$ & $1.00151 - 1.00199$ & $5.87 - 6.20$\\
\hline
  \textbf{Ep3}  & $<0.29$             &                    & $<0.27$              &               & $0.13 - 0.26$       &  \\
                & $0.99978 - 1.00006$ &                    & $0.99980 - 1.00002$  &               & $0.99981 - 0.99992$ & \\
                & $1.00020 - 1.00076$ & $6.28 - 8.85$      & $1.00042 - 1.00074$  & $6.34 - 7.34$ & $1.00059 - 1.00073$ & $6.35 - 6.72$\\
\hline
  \textbf{Ep4}  & $0.60490 - 1.00818$ &  $3.68 - 5.23$     & $0.84716 - 1.00337$  &               & $0.93259 - 1.00233$ &  \\
                &                     &                    & $1.00474 - 1.00794$  & $3.71 - 4.32$ & $1.00647 - 1.00788$ & $3.72 - 3.94$\\
\hline
  \textbf{Ep5}  & $0.86756 - 1.03363$ &  $3.05 - 4.34$     & $0.99257-1.03363$    & $3.08 - 3.59$ & $1.01310 - 1.01590$ &  \\
                &                     &                    &                      &               & $1.02148 - 1.03363$ & $3.08 - 3.27$\\
\hline
  \textbf{Kep1} & $<0.55$             & $4.17 - 5.99$      & $0.29 - 0.54$        & $4.21 - 4.92$ & $0.45 - 0.53$       & $4.22 - 4.48$\\
\hline
  \textbf{Kep2} & $<0.44$             & $5.17 - 7.34$      & $0.12 - 0.43$        & $5.22 - 6.06$ & $0.31 - 0.42$       & $5.23 - 5.54$\\
\hline
  \textbf{Kep3} & $<0.24$             & $4.75 - 6.81$      & $<0.23$              & $4.79 - 5.59$ & $0.11 - 0.22$       & $4.80 - 5.09$\\
\hline
  \textbf{Kep4} & $0.57 - 0.93$       & $2.98 - 4.33$      & $0.80 - 0.93$        & $3.01 - 3.54$ & $0.88 - 0.92$       & $3.02 - 3.21$\\
\hline
 \textbf{Kep5}  & $0.95615 - 1.18902$ & $2.84 - 4.10$      & $1.11049 - 1.18574$  & $2.87 - 3.36$ & $1.16028 - 1.18489$ & $2.87 - 3.05$\\
\hline \hline
\end{tabular}
\end{center}
\end{table*}
%%%%%%%%%%%%%%%%%%%%

Predictions of these ten newly considered models are illustrated in Figs.~\ref{figure:Keplerian}--\ref{figure:nuEpicyclic}. The~observationally determined ranges of $M\times\nuU (a)$ for each of the~Galactic microquasars are denoted in Figs.~\ref{figure:Keplerian}b and \ref{figure:nuEpicyclic} by the~individual horizontal grey-scaled areas. For a~particular QPO model and source with an independently estimated mass range we calculated the~relevant spin interval of the~central compact object as predicted by the~concrete model. The~results are summarized in Table \ref{table:spin} and show that there could be two (for the~Kep models) or up to four (for the~Ep models) different spin intervals for each QPO model. The~number of the~relevant spin intervals and their ranges depend on the~topology of the~$M\times\nuU (a)$ relation and on the~observed range of mass for a~given source. Confronting the~QPO models predictions with the~data of Galactic microquasars\footnote{The~assumed values of the~measured mass and QPO frequencies of each microquasar are summarized in Table 1 of Paper~I, and follow from the~studies of \citet{gre-etal:2001,grn-etal:2001,str:2001,oro-etal:2002,rem-etal:2002,mcc-rem:2006}.}, we find that each of these models is compatible with the~assumption of $a>1$, but only some of them with the~assumption of $a\gtrsim 1$.

Comparing our new results with the~results obtained in Paper~I, we conclude that five Keplerian forced resonance models (Kep1--Kep5) show qualitatively similar behaviour with the~3:2 parametric Keplerian model (Kep) considered in Paper~I. More specifically, for all the~Kep models we find the~existence of a~unique QPO excitation radius $\tilde{r}_{\mathrm{p:q}}$ and consequent simple behaviour of $M\times\nuU (a)$ relations represented by a~single function having solely one maximum close to $a\gtrsim1$. Inspecting the~resonant frequencies drawn in Fig.~\ref{figure:Keplerian}b and the~results given in Table~\ref{table:spin} we can find that only one of the~Keplerian forced resonance models (Kep5) is compatible with the~expectation of $a\gtrsim 1$ and gives quantitatively similar results to the~3:2 parametric Keplerian model (Kep) considered in Paper~I. For the~other four Kep models (Kep1--Kep4), the~inferred values of spin are either lower or several times higher than the~extreme Kerr black hole value $a = 1$. Therefore, for the~case of the~three Galactic microquasars considered here, these four Kep models can be ruled out from further consideration.

On the~other hand, from our results given in Fig.~\ref{figure:nuEpicyclic} and Table~\ref{table:spin}, we can see that each of the~five epicyclic forced resonance models (Ep1--Ep5) is compatible with the~expectation of $a\gtrsim 1$, and thus none of them can be ruled out. The~Ep models imply the~existence of multiple resonant radii and therefore more complicated $M\times\nuU (a)$ relations than those implied by the~Kep models. The~Ep1--Ep5 models are similar to the~3:2 parametric epicyclic model (Ep) considered in Paper~I. However, as elaborated in Sect.~\ref{section:topologies}, the~Ep models in general allow a~very wide variety of topologies of the~$M\times\nuU (a)$ relation. The~relation explored in Paper~I for the~Ep model corresponds to only one specific topology from this variety. For the~Ep models considered here, three different topologies are possible and for only one of them (Ep5) the~$M\times\nuU (a)$ relation has the~same topology as that explored in Paper~I.

For all the~Ep models, there is the~possibility of recognizing a~direct observational signature of presence of a~super-spinning compact object. It is clear from Figs.~\ref{figure:nuEpicyclic}b--\ref{figure:nuEpicyclic}d that more pairs of different 3:2 commensurable frequencies can be expected within a~single source. In a~full analogy to the~specific case of Ep model discussed in Paper~I, we conclude that this issue can be resolved using the~large amount of high-quality data available from the~next generation of X-ray observatories including technologies such as the~proposed concept of Large Area Detector \citep[][]{fer-etal:2012,fer-etal:2014:,Zha-etal:2016:}.

\subsection*{Spin intervals implied by 3:2 or 2:3 QPOs}

The~results presented in Table \ref{table:spin} assume that the~predicted pair of QPO frequencies forms the~3:2 ratio, whereas the~frequency denoted in Table \ref{table:models} as $\nuU$  is higher than that denoted as $\nuL$. However, some of the~considered models imply that there can be an inverse inequality, $\nuL/\nuU=3/2$. We have investigated this possibility and found that four models are compatible with observation. Specifically, for the~Kep4 model the~implied spin intervals are then the~same as those given in Table \ref{table:spin} for the~Kep5 model and vice versa. A~fully analogous conclusion also holds for the~Ep4 and Ep5 models.

%-------------------------------------------------------------------------------
\section*{Acknowledgments}
%-------------------------------------------------------------------------------
{We would like to acknowledge the~Czech grant GA\v{C}R 17-16287S, `Oscillations and coherent features in black-hole accretion disks and their observational signatures'. We thank to the~INTER-EXCELLENCE project No. LTI17018 which supports international activities of the~Silesian University in Opava (SU) and Astronomical Institute in Prague (ASU). We also acknowledge two internal SU grants, SGS/14,15/2016 and IGS/2/2016. ZS acknowledges the~Albert Einstein Center for Gravitation and Astrophysics supported by the~Czech Science Foundation grant No. 14-37086G. We are grateful to Marek Abramowicz (Nicolaus Copernicus Astronomical Center in Warsaw -- CAMK \& SU), Wlodek Klu{\'z}niak (CAMK), {Macziek Wilgus} (CAMK), Ji\v{r}\'{i} Hor\'{a}k (ASU), and Pavel Bakala (SU) for many useful discussions. Furthermore we would like to acknowledge the~hospitality of CAMK. We express our sincere thanks to the~concierges of the~Ml\'{y}nsk\'{a} hotel in Uhersk\'{e} Hradi\v{s}t\v{e}, Czech Republic for their kind help and participation in organizing frequent workshops of SU and ASU. Last but not least, the~authors wish to thank the~anonymous referee for the~useful and constructive comments that substantially helped to improve the~manuscript.}

%\clearpage

\bibliographystyle{aa}
\bibliography{reference-2017}

\begin{thebibliography}{90}
\expandafter\ifx\csname natexlab\endcsname\relax\def\natexlab#1{#1}\fi

\bibitem[{{Abramowicz} {et~al.}(2002){Abramowicz}, {Almergren}, {Klu{\'z}niak},
  {Thampan}, \& {Wallinder}}]{abr-etal:2002:}
{Abramowicz}, M.~A., {Almergren}, G.~J.~E., {Klu{\'z}niak}, W., {Thampan},
  A.~V., \& {Wallinder}, F. 2002, CQG, 19, L57

\bibitem[{{Abramowicz} {et~al.}(2003{\natexlab{a}}){Abramowicz}, {Bulik},
  {Bursa}, \& {Klu{\'z}niak}}]{abr-etal:2003b}
{Abramowicz}, M.~A., {Bulik}, T., {Bursa}, M., \& {Klu{\'z}niak}, W.
  2003{\natexlab{a}}, A\&A, 404, L21

\bibitem[{{Abramowicz} {et~al.}(2003{\natexlab{b}}){Abramowicz}, {Karas},
  {Klu{\'z}niak}, {Lee}, \& {Rebusco}}]{abr-etal:2003c}
{Abramowicz}, M.~A., {Karas}, V., {Klu{\'z}niak}, W., {Lee}, W.~H., \&
  {Rebusco}, P. 2003{\natexlab{b}}, PASJ, 55, 467

\bibitem[{{Abramowicz} \& {Klu{\'z}niak}(2001)}]{abr-klu:2001}
{Abramowicz}, M.~A. \& {Klu{\'z}niak}, W. 2001, A\&A, 374, L19

\bibitem[{{Abramowicz} {et~al.}(2004){Abramowicz}, {Klu{\'z}niak},
  {McClintock}, \& {Remillard}}]{abr-etal:2004:ApJ:}
{Abramowicz}, M.~A., {Klu{\'z}niak}, W., {McClintock}, J.~E., \& {Remillard},
  R.~A. 2004, ApJ, 609, L63

\bibitem[{{Aliev} \& {Galtsov}(1981)}]{ali-gal:1981}
{Aliev}, A.~N. \& {Galtsov}, D.~V. 1981, GRG, 13, 899

\bibitem[{{Aschenbach} {et~al.}(2004){Aschenbach}, {Grosso}, {Porquet}, \&
  {Predehl}}]{asc-etal:2004}
{Aschenbach}, B., {Grosso}, N., {Porquet}, D., \& {Predehl}, P. 2004, A\&A,
  417, 71

\bibitem[{{Bambi}(2011)}]{bam:2011}
{Bambi}, C. 2011, EPL (Europhysics Letters), 94, 50002

\bibitem[{{Bambi}(2012)}]{bam:2012}
{Bambi}, C. 2012, Journal of Cosmology and Astroparticle Physics, 9, 14

\bibitem[{{Bambi}(2014)}]{bam:2014}
{Bambi}, C. 2014, Physics Letters B, 730, 59

\bibitem[{{Bambi} \& {Freese}(2009)}]{bam-fre:2009}
{Bambi}, C. \& {Freese}, K. 2009, PRD, 79, 043002

\bibitem[{{Calvani} \& {Nobili}(1979)}]{cal-nob:1979}
{Calvani}, M. \& {Nobili}, L. 1979, Nuovo Cimento B Serie, 51, 247

\bibitem[{{Done} {et~al.}(2007){Done}, {Gierli{\'n}ski}, \&
  {Kubota}}]{don-etal:2007}
{Done}, C., {Gierli{\'n}ski}, M., \& {Kubota}, A. 2007, The Astronomy and
  Astrophysics Review, 15, 1

\bibitem[{{Feroci} {et~al.}(2014){Feroci}, {den Herder}, {Bozzo}, {Barret},
  {Brandt}, {Hernanz}, {van der Klis}, {Pohl}, {Santangelo}, {Stella}, \&
  et~al.}]{fer-etal:2014:}
{Feroci}, M., {den Herder}, J.~W., {Bozzo}, E., {et~al.} 2014, in \procspie,
  Vol. 9144, Space Telescopes and Instrumentation 2014: Ultraviolet to Gamma
  Ray, 91442T

\bibitem[{{Feroci} {et~al.}(2012){Feroci}, {Stella}, {van der Klis},
  {Courvoisier}, {Hernanz}, {Hudec}, {Santangelo}, {Walton}, {Zdziarski},
  {Barret}, {Belloni}, {Braga}, {Brandt}, {Budtz-J{\o}rgensen}, {Campana}, {den
  Herder}, {Huovelin}, {Israel}, {Pohl}, {Ray}, {Vacchi}, {Zane}, {Argan},
  {Attin{\`a}}, {Bertuccio}, {Bozzo}, {Campana}, {Chakrabarty}, {Costa}, {De
  Rosa}, {Del Monte}, {Di Cosimo}, {Donnarumma}, {Evangelista}, {Haas},
  {Jonker}, {Korpela}, {Labanti}, {Malcovati}, {Mignani}, {Muleri},
  {Rapisarda}, {Rashevsky}, {Rea}, {Rubini}, {Tenzer}, {Wilson-Hodge},
  {Winter}, {Wood}, {Zampa}, {Zampa}, {Abramowicz}, {Alpar}, {Altamirano},
  {Alvarez}, {Amati}, {Amoros}, {Antonelli}, {Artigue}, {Azzarello},
  {Bachetti}, {Baldazzi}, {Barbera}, {Barbieri}, {Basa}, {Baykal}, {Belmont},
  {Boirin}, {Bonvicini}, {Burderi}, {Bursa}, {Cabanac}, {Cackett}, {Caliandro},
  {Casella}, {Chaty}, {Chenevez}, {Coe}, {Collura}, {Corongiu}, {Covino},
  {Cusumano}, {D'Amico}, {Dall'Osso}, {De Martino}, {De Paris}, {Di Persio},
  {Di Salvo}, {Done}, {Dov{\v c}iak}, {Drago}, {Ertan}, {Fabiani}, {Falanga},
  {Fender}, {Ferrando}, {Della Monica Ferreira}, {Fraser}, {Frontera},
  {Fuschino}, {Galvez}, {Gandhi}, {Giommi}, {Godet}, {G{\"o}{\v g}{\"u}{\c s}},
  {Goldwurm}, {G{\"o}tz}, {Grassi}, {Guttridge}, {Hakala}, {Henri}, {Hermsen},
  {Horak}, {Hornstrup}, {in't Zand}, {Isern}, {Kalemci}, {Kanbach}, {Karas},
  {Kataria}, {Kennedy}, {Klochkov}, {Klu{\'z}niak}, {Kokkotas}, {Kreykenbohm},
  {Krolik}, {Kuiper}, {Kuvvetli}, {Kylafis}, {Lattimer}, {Lazzarotto}, {Leahy},
  {Lebrun}, {Lin}, {Lund}, {Maccarone}, {Malzac}, {Marisaldi}, {Martindale},
  {Mastropietro}, {McClintock}, {McHardy}, {Mendez}, {Mereghetti}, {Miller},
  {Mineo}, {Morelli}, {Morsink}, {Motch}, {Motta}, {Mu{\~n}oz-Darias},
  {Naletto}, {Neustroev}, {Nevalainen}, {Olive}, {Orio}, {Orlandini},
  {Orleanski}, {Ozel}, {Pacciani}, {Paltani}, {Papadakis}, {Papitto},
  {Patruno}, {Pellizzoni}, {Petr{\'a}{\v c}ek}, {Petri}, {Petrucci}, {Phlips},
  {Picolli}, {Possenti}, {Psaltis}, {Rambaud}, {Reig}, {Remillard},
  {Rodriguez}, {Romano}, {Romanova}, {Schanz}, {Schmid}, {Segreto}, {Shearer},
  {Smith}, {Smith}, {Soffitta}, {Stergioulas}, {Stolarski}, {Stuchlik},
  {Tiengo}, {Torres}, {T{\"o}r{\"o}k}, {Turolla}, {Uttley}, {Vaughan},
  {Vercellone}, {Waters}, {Watts}, {Wawrzaszek}, {Webb}, {Wilms}, {Zampieri},
  {Zezas}, \& {Ziolkowski}}]{fer-etal:2012}
{Feroci}, M., {Stella}, L., {van der Klis}, M., {et~al.} 2012, Experimental
  Astronomy, 34, 415

\bibitem[{{Gilfanov} {et~al.}(2000){Gilfanov}, {Churazov}, \&
  {Revnivtsev}}]{gilf-etal:2000}
{Gilfanov}, M., {Churazov}, E., \& {Revnivtsev}, M. 2000, MNRAS, 316, 923

\bibitem[{{Gimon} \& {Ho{\v r}ava}(2009)}]{gim-hor:2009}
{Gimon}, E.~G. \& {Ho{\v r}ava}, P. 2009, Physics Letters B, 672, 299

\bibitem[{{Greene} {et~al.}(2001){Greene}, {Bailyn}, \&
  {Orosz}}]{gre-etal:2001}
{Greene}, J., {Bailyn}, C.~D., \& {Orosz}, J.~A. 2001, ApJ, 554, 1290

\bibitem[{{Greiner} {et~al.}(2001){Greiner}, {Cuby}, \&
  {McCaughrean}}]{grn-etal:2001}
{Greiner}, J., {Cuby}, J.~G., \& {McCaughrean}, M.~J. 2001, Nature, 414, 522

\bibitem[{{Hor{\'a}k}(2004)}]{hor:2004}
{Hor{\'a}k}, J. 2004, in RAGtime 4/5: Workshops on black holes and neutron
  stars, ed. S.~{Hled{\'{\i}}k} \& Z.~{Stuchl{\'{\i}}k}, 91--110

\bibitem[{{Hor{\'a}k}(2005)}]{hor:2005b}
{Hor{\'a}k}, J. 2005, AN, 326, 824

\bibitem[{{Hor{\'a}k}(2008)}]{hor:2008}
{Hor{\'a}k}, J. 2008, A\&A, 486, 1

\bibitem[{{Hor{\'a}k} {et~al.}(2009){Hor{\'a}k}, {Abramowicz}, {Klu{\'z}niak},
  {Rebusco}, \& {T{\"o}r{\"o}k}}]{hor-etal:2009}
{Hor{\'a}k}, J., {Abramowicz}, M.~A., {Klu{\'z}niak}, W., {Rebusco}, P., \&
  {T{\"o}r{\"o}k}, G. 2009, A\&A, 499, 535

\bibitem[{{Hor{\'a}k} \& {Karas}(2006)}]{hor-kar:2006}
{Hor{\'a}k}, J. \& {Karas}, V. 2006, A\&A, 451, 377

\bibitem[{{Hor{\'a}k} \& {Lai}(2013)}]{hor-don:2014}
{Hor{\'a}k}, J. \& {Lai}, D. 2013, MNRAS, 434, 2761

\bibitem[{{Ingram} \& {Done}(2010)}]{Ing-Don:2010:}
{Ingram}, A. \& {Done}, C. 2010, MNRAS, 405, 2447

\bibitem[{{Ingram} \& {Done}(2011)}]{Ing-Don:2011:}
{Ingram}, A. \& {Done}, C. 2011, MNRAS, 415, 2323

\bibitem[{{Ingram} {et~al.}(2009){Ingram}, {Done}, \&
  {Fragile}}]{Ing-Don:2009:}
{Ingram}, A., {Done}, C., \& {Fragile}, P.~C. 2009, MNRAS, 397, L101

\bibitem[{{Ingram} {et~al.}(2016){Ingram}, {van der Klis}, {Middleton}, {Done},
  {Altamirano}, {Heil}, {Uttley}, \& {Axelsson}}]{Ing-Don:2016:}
{Ingram}, A., {van der Klis}, M., {Middleton}, M., {et~al.} 2016, MNRAS, 461,
  1967

\bibitem[{{Johannsen}(2016)}]{joh:2016}
{Johannsen}, T. 2016, CQG, 33, 124001

\bibitem[{{Johannsen} \& {Psaltis}(2011)}]{joh-psa:2011}
{Johannsen}, T. \& {Psaltis}, D. 2011, ApJ, 726, 11

\bibitem[{{Kato}(2004)}]{kat:2004}
{Kato}, S. 2004, PASJ, 56, L25

\bibitem[{{Klu{\'z}niak} \& {Abramowicz}(2001)}]{klu-abr:2001}
{Klu{\'z}niak}, W. \& {Abramowicz}, M.~A. 2001, ArXiv e-prints
  [\eprint[arXiv]{astro-ph/0105057}]

\bibitem[{{Klu{\'z}niak} \& {Abramowicz}(2002)}]{klu-abr:2002:}
{Klu{\'z}niak}, W. \& {Abramowicz}, M.~A. 2002, ArXiv e-prints
  [\eprint[arXiv]{astro-ph/0203314}]

\bibitem[{{Klu{\'z}niak} \& {Abramowicz}(2005)}]{klu-abr:2005:}
{Klu{\'z}niak}, W. \& {Abramowicz}, M.~A. 2005, Ap\&SS, 300, 143

\bibitem[{{Kolo{\v s}} \& {Stuchl{\'{i}}k}(2013)}]{kol-stu:2013}
{Kolo{\v s}}, M. \& {Stuchl{\'{i}}k}, Z. 2013, PRD, 88, 065004

\bibitem[{{Kotrlov{\'a}} {et~al.}(2008){Kotrlov{\'a}}, {Stuchl{\'{i}}k}, \&
  {T{\"o}r{\"o}k}}]{kot-etal:2008:CQG:}
{Kotrlov{\'a}}, A., {Stuchl{\'{i}}k}, Z., \& {T{\"o}r{\"o}k}, G. 2008, CQG, 25,
  225016

\bibitem[{{Kotrlov{\'a}} {et~al.}(2014){Kotrlov{\'a}}, {T{\"o}r{\"o}k}, {{\v
  S}r{\'a}mkov{\'a}}, \& {Stuchl{\'{i}}k}}]{kot-etal:2014:AA:}
{Kotrlov{\'a}}, A., {T{\"o}r{\"o}k}, G., {{\v S}r{\'a}mkov{\'a}}, E., \&
  {Stuchl{\'{i}}k}, Z. 2014, A\&A, 572, A79

\bibitem[{{Li} \& {Bambi}(2013{\natexlab{a}})}]{li-bam:2013b}
{Li}, Z. \& {Bambi}, C. 2013{\natexlab{a}}, PRD, 87, 124022

\bibitem[{{Li} \& {Bambi}(2013{\natexlab{b}})}]{li-bam:2013a}
{Li}, Z. \& {Bambi}, C. 2013{\natexlab{b}}, Journal of Cosmology and
  Astroparticle Physics, 3, 31

\bibitem[{{McClintock} {et~al.}(2011){McClintock}, {Narayan}, {Davis}, {Gou},
  {Kulkarni}, {Orosz}, {Penna}, {Remillard}, \& {Steiner}}]{mcc-etal:2011}
{McClintock}, J.~E., {Narayan}, R., {Davis}, S.~W., {et~al.} 2011, CQG, 28,
  114009

\bibitem[{{McClintock} {et~al.}(2010){McClintock}, {Narayan}, {Gou}, {Liu},
  {Penna}, \& {Steiner}}]{mcc-etal:2010}
{McClintock}, J.~E., {Narayan}, R., {Gou}, L., {et~al.} 2010, X-ray Astronomy
  2009; Present Status, Multi-Wavelength Approach and Future Perspectives,
  1248, 101

\bibitem[{{McClintock} {et~al.}(2007){McClintock}, {Narayan}, \&
  {Shafee}}]{mcc-etal:2008}
{McClintock}, J.~E., {Narayan}, R., \& {Shafee}, R. 2007, ArXiv e-prints
  [\eprint[arXiv]{0707.4492}]

\bibitem[{{McClintock} {et~al.}(2014){McClintock}, {Narayan}, \&
  {Steiner}}]{mcc-etal:2014}
{McClintock}, J.~E., {Narayan}, R., \& {Steiner}, J.~F. 2014, Space Sci. Rev.,
  183, 295

\bibitem[{{McClintock} \& {Remillard}(2006)}]{mcc-rem:2006}
{McClintock}, J.~E. \& {Remillard}, R.~A. 2006, {Black hole binaries} (In:
  Compact stellar X-ray sources. Edited by Walter Lewin \& Michiel van der
  Klis. Cambridge Astrophysics Series, No. 39. Cambridge, UK: Cambridge
  University Press), 157--213

\bibitem[{{Middleton} {et~al.}(2006){Middleton}, {Del Pozzo}, {Farr}, {Sesana},
  \& {Vecchio}}]{mid-etal:2006}
{Middleton}, H., {Del Pozzo}, W., {Farr}, W.~M., {Sesana}, A., \& {Vecchio}, A.
  2006, MNRAS, 455, L72

\bibitem[{{Miller}(2007)}]{mil:2007}
{Miller}, J.~M. 2007, ARA\&A, 45, 441

\bibitem[{{Montero} {et~al.}(2007){Montero}, {Zanotti}, {Font}, \&
  {Rezzolla}}]{Montero-etal:2007:}
{Montero}, P.~J., {Zanotti}, O., {Font}, J.~A., \& {Rezzolla}, L. 2007, MNRAS,
  378, 1101

\bibitem[{{Motta} {et~al.}(2014){Motta}, {Mu{\~n}oz-Darias}, {Sanna}, {Fender},
  {Belloni}, \& {Stella}}]{mot-etal:2014}
{Motta}, S.~E., {Mu{\~n}oz-Darias}, T., {Sanna}, A., {et~al.} 2014, MNRAS, 439,
  L65

\bibitem[{{Nowak} \& {Lehr}(1999)}]{now-leh:1999}
{Nowak}, M. \& {Lehr}, D. 1999, in Theory of Black Hole Accretion Disks, ed.
  M.~A. {Abramowicz}, G.~{Bj{\"o}rnsson}, \& J.~E. {Pringle} (Cambridge:
  Cambridge University Press), 21

\bibitem[{{Orosz} {et~al.}(2002){Orosz}, {Groot}, {van der Klis}, {McClintock},
  {Garcia}, {Zhao}, {Jain}, {Bailyn}, \& {Remillard}}]{oro-etal:2002}
{Orosz}, J.~A., {Groot}, P.~J., {van der Klis}, M., {et~al.} 2002, ApJ, 568,
  845

\bibitem[{{Ortega-Rodr{\'{\i}}guez} {et~al.}(2014){Ortega-Rodr{\'{\i}}guez},
  {Sol{\'{\i}}s-S{\'a}nchez}, {L{\'o}pez-Barquero}, {Matamoros-Alvarado}, \&
  {Venegas-Li}}]{ort-etal:2014}
{Ortega-Rodr{\'{\i}}guez}, M., {Sol{\'{\i}}s-S{\'a}nchez}, H.,
  {L{\'o}pez-Barquero}, V., {Matamoros-Alvarado}, B., \& {Venegas-Li}, A. 2014,
  MNRAS, 440, 3011

\bibitem[{{Psaltis} {et~al.}(2008){Psaltis}, {Perrodin}, {Dienes}, \&
  {Mocioiu}}]{psa-etal:2008}
{Psaltis}, D., {Perrodin}, D., {Dienes}, K.~R., \& {Mocioiu}, I. 2008, PRL,
  100, 091101

\bibitem[{{Rebusco}(2004)}]{reb:2004}
{Rebusco}, P. 2004, PASJ, 56, 553

\bibitem[{{Rebusco}(2008)}]{reb:2008}
{Rebusco}, P. 2008, New Astronomy Reviews, 51, 855

\bibitem[{{Remillard} {et~al.}(2002){Remillard}, {Muno}, {McClintock}, \&
  {Orosz}}]{rem-etal:2002}
{Remillard}, R.~A., {Muno}, M.~P., {McClintock}, J.~E., \& {Orosz}, J.~A. 2002,
  ApJ, 580, 1030

\bibitem[{{Rezzolla} {et~al.}(2003){Rezzolla}, {Yoshida}, \&
  {Zanotti}}]{rez-etal:2003}
{Rezzolla}, L., {Yoshida}, S., \& {Zanotti}, O. 2003, MNRAS, 344, 978

\bibitem[{{Schnittman} \& {Rezzolla}(2006)}]{sch-rez:2006:}
{Schnittman}, J.~D. \& {Rezzolla}, L. 2006, ApJ, 637, L113

\bibitem[{{Shafee} {et~al.}(2008){Shafee}, {McKinney}, {Narayan},
  {Tchekhovskoy}, {Gammie}, \& {McClintock}}]{sha-etal:2008}
{Shafee}, R., {McKinney}, J.~C., {Narayan}, R., {et~al.} 2008, ApJ, 687, L25

\bibitem[{{\v{S}r{\'a}mkov{\'a}} {et~al.}(2007){\v{S}r{\'a}mkov{\'a}},
  {Torkelsson}, \& {Abramowicz}}]{sra-etal:2007}
{\v{S}r{\'a}mkov{\'a}}, E., {Torkelsson}, U., \& {Abramowicz}, M.~A. 2007,
  A\&A, 467, 641

\bibitem[{{Steiner} {et~al.}(2011){Steiner}, { Reis}, {McClintock}, {Narayan},
  {Remillard}, {Orosz}, {Gou}, {Fabian}, \& {Torres}}]{ste-etal:2011}
{Steiner}, J.~F., { Reis}, R.~C., {McClintock}, J.~E., {et~al.} 2011, MNRAS,
  416, 941

\bibitem[{{Strohmayer}(2001)}]{str:2001}
{Strohmayer}, T.~E. 2001, ApJ, 552, L49

\bibitem[{{Stuchl\'{i}k}(1980)}]{stu:1980}
{Stuchl\'{i}k}, Z. 1980, Bulletin of the Astronomical Institutes of
  Czechoslovakia, 31, 129

\bibitem[{{Stuchl\'{i}k}(1981)}]{stu:1981}
{Stuchl\'{i}k}, Z. 1981, Bulletin of the Astronomical Institutes of
  Czechoslovakia, 32, 68

\bibitem[{{Stuchl{\'{\i}}k} {et~al.}(2011){Stuchl{\'{\i}}k}, {Hled{\'{\i}}k},
  \& {Truparov{\'a}}}]{stu-etal:2011}
{Stuchl{\'{\i}}k}, Z., {Hled{\'{\i}}k}, S., \& {Truparov{\'a}}, K. 2011, CQG,
  28, 155017

\bibitem[{{Stuchl{\'{\i}}k} \& {Kolo{\v
  s}}(2016{\natexlab{a}})}]{stu-kol:2016:ApJ}
{Stuchl{\'{\i}}k}, Z. \& {Kolo{\v s}}, M. 2016{\natexlab{a}}, APJ, 825, 13

\bibitem[{{Stuchl{\'{\i}}k} \& {Kolo{\v
  s}}(2016{\natexlab{b}})}]{stu-kol:2016:AA}
{Stuchl{\'{\i}}k}, Z. \& {Kolo{\v s}}, M. 2016{\natexlab{b}}, \aap, 586, A130

\bibitem[{{Stuchl{\'{i}}k} \& {Kotrlov{\'a}}(2009)}]{stu-kot:2009}
{Stuchl{\'{i}}k}, Z. \& {Kotrlov{\'a}}, A. 2009, GRG, 41, 1305

\bibitem[{{Stuchl{\'{\i}}k} {et~al.}(2008){Stuchl{\'{\i}}k}, {Kotrlov{\'a}}, \&
  {G.~T{\"o}r{\"o}k}}]{stu-etal:2008b}
{Stuchl{\'{\i}}k}, Z., {Kotrlov{\'a}}, A., \& {G.~T{\"o}r{\"o}k}. 2008, Acta
  Astron., 58, 441

\bibitem[{{Stuchl{\'{i}}k} {et~al.}(2013){Stuchl{\'{i}}k}, {Kotrlov{\'a}}, \&
  {T{\"o}r{\"o}k}}]{stu-etal:2013:AA:}
{Stuchl{\'{i}}k}, Z., {Kotrlov{\'a}}, A., \& {T{\"o}r{\"o}k}, G. 2013, A\&A,
  552, A10

\bibitem[{{Stuchl{\'{i}}k} {et~al.}(2014){Stuchl{\'{i}}k}, {Kotrlov\'{a}},
  {T\"{o}r\"{o}k}, \& {Goluchov\'{a}}}]{stu-etal:2014:ACTA:}
{Stuchl{\'{i}}k}, Z., {Kotrlov\'{a}}, A., {T\"{o}r\"{o}k}, G., \&
  {Goluchov\'{a}}, K. 2014, Acta Astron., 64, 45

\bibitem[{{Stuchl{\'{i}}k} \& {Schee}(2010)}]{stu-sch:2010}
{Stuchl{\'{i}}k}, Z. \& {Schee}, J. 2010, CQG, 27, 215017

\bibitem[{{Stuchl{\'{i}}k} \& {Schee}(2012{\natexlab{a}})}]{stu-sch:2012b}
{Stuchl{\'{i}}k}, Z. \& {Schee}, J. 2012{\natexlab{a}}, CQG, 29, 025008

\bibitem[{{Stuchl{\'{i}}k} \& {Schee}(2012{\natexlab{b}})}]{stu-sch:2012a}
{Stuchl{\'{i}}k}, Z. \& {Schee}, J. 2012{\natexlab{b}}, CQG, 29, 065002

\bibitem[{{Stuchl{\'{i}}k} \& {Schee}(2013)}]{stu-sch:2013}
{Stuchl{\'{i}}k}, Z. \& {Schee}, J. 2013, CQG, 30, 075012

\bibitem[{{T{\"o}r{\"o}k}(2005{\natexlab{a}})}]{tor:2005:AA:}
{T{\"o}r{\"o}k}, G. 2005{\natexlab{a}}, A\&A, 440, 1

\bibitem[{{T{\"o}r{\"o}k}(2005{\natexlab{b}})}]{tor:2005:AN:}
{T{\"o}r{\"o}k}, G. 2005{\natexlab{b}}, AN, 326, 856

\bibitem[{{T{\"o}r{\"o}k} {et~al.}(2005){T{\"o}r{\"o}k}, {Abramowicz},
  {Klu{\'z}niak}, \& {Stuchl{\'{i}}k}}]{tor-etal:2005:AA:}
{T{\"o}r{\"o}k}, G., {Abramowicz}, M.~A., {Klu{\'z}niak}, W., \&
  {Stuchl{\'{i}}k}, Z. 2005, A\&A, 436, 1

\bibitem[{{T{\"o}r{\"o}k} {et~al.}(2016){T{\"o}r{\"o}k}, {Goluchov{\'a}},
  {Hor{\'a}k}, {{\v S}r{\'a}mkov{\'a}}, {Urbanec}, {Pech{\'a}{\v c}ek}, \&
  {Bakala}}]{tor-etal:2016:MNRAS:}
{T{\"o}r{\"o}k}, G., {Goluchov{\'a}}, K., {Hor{\'a}k}, J., {et~al.} 2016,
  MNRAS, 457, L19

\bibitem[{{T{\"o}r{\"o}k} {et~al.}(2011){T{\"o}r{\"o}k}, {Kotrlov{\'a}}, {{\v
  S}r{\'a}mkov{\'a}}, \& {Stuchl{\'{i}}k}}]{tor-etal:2011:AA:}
{T{\"o}r{\"o}k}, G., {Kotrlov{\'a}}, A., {{\v S}r{\'a}mkov{\'a}}, E., \&
  {Stuchl{\'{i}}k}, Z. 2011, A\&A, 531, A59

\bibitem[{{T{\"o}r{\"o}k} \& {Stuchl{\'{i}}k}(2005)}]{tor-stu:2005:AA:}
{T{\"o}r{\"o}k}, G. \& {Stuchl{\'{i}}k}, Z. 2005, A\&A, 437, 775

\bibitem[{{{\v S}r{\'a}mkov{\'a}}(2005)}]{sra:2005:}
{{\v S}r{\'a}mkov{\'a}}, E. 2005, Astronomische Nachrichten, 326, 835

\bibitem[{{van der Klis}(2006)}]{kli:2006}
{van der Klis}, M. 2006, {Rapid X-ray Variability} (In: Compact stellar X-ray
  sources. Edited by Walter Lewin \& Michiel van der Klis. Cambridge
  Astrophysics Series, No. 39. Cambridge, UK: Cambridge University Press),
  39--112

\bibitem[{{Vio} {et~al.}(2006){Vio}, {Rebusco}, {Andreani}, {Madsen}, \&
  {Overgaard}}]{vio-etal:2006}
{Vio}, R., {Rebusco}, P., {Andreani}, P., {Madsen}, H., \& {Overgaard}, R.~V.
  2006, A\&A, 452, 383

\bibitem[{{Wagoner}(2012)}]{wag:2012}
{Wagoner}, R.~V. 2012, APJL, 752, L18

\bibitem[{{Wagoner} {et~al.}(2001){Wagoner}, {Silbergleit}, \&
  {Ortega-Rodr{\'{\i}}guez}}]{wag-etal:2001}
{Wagoner}, R.~V., {Silbergleit}, A.~S., \& {Ortega-Rodr{\'{\i}}guez}, M. 2001,
  APJL, 559, L25

\bibitem[{{Yagi} \& {Stein}(2016)}]{yag-ste:2016:}
{Yagi}, K. \& {Stein}, L.~C. 2016, Classical and Quantum Gravity, 33, 054001

\bibitem[{{Zanotti} {et~al.}(2005){Zanotti}, {Font}, {Rezzolla}, \&
  {Montero}}]{zan-etal:2005:}
{Zanotti}, O., {Font}, J.~A., {Rezzolla}, L., \& {Montero}, P.~J. 2005, MNRAS,
  356, 1371

\bibitem[{{Zhang} {et~al.}(2016){Zhang}, {Feroci}, {Santangelo}, {Dong},
  {Feng}, {Lu}, {Nandra}, {Wang}, {Zhang}, {Bozzo}, {Brandt}, {De Rosa}, {Gou},
  {Hernanz}, {van der Klis}, {Li}, {Liu}, {Orleanski}, {Pareschi}, {Pohl},
  {Poutanen}, {Qu}, {Schanne}, {Stella}, {Uttley}, \& {Watts}}]{Zha-etal:2016:}
{Zhang}, S.~N., {Feroci}, M., {Santangelo}, A., {et~al.} 2016, in \procspie,
  Vol. 9905, Society of Photo-Optical Instrumentation Engineers (SPIE)
  Conference Series, 99051Q

\bibitem[{{Zhou} {et~al.}(2015){Zhou}, {Yuan}, {Pan}, \& {Liu}}]{zho-etal:2015}
{Zhou}, X.-L., {Yuan}, W., {Pan}, H.-W., \& {Liu}, Z. 2015, APJL, 798, L5

\end{thebibliography}

\end{document}